# Interaction between Mo and intrinsic or extrinsic defects of Mo doped LiNbO₃ from first-principles calculations

Weiwei Wang,[1] Hongde Liu,[1,a)] Dahuai Zheng,[1] Yongfa Kong,[1,b)] Lixin Zhang,[1] and Jingjun Xu,[1]

[1]*MOE Key Laboratory of Weak-Light Nonlinear Photonics, School of Physics and TEDA Institute of Applied Physics, Nankai University, Tianjin 300071, China*

a) E-mail: liuhd97@nankai.edu.cn (H. Liu); b) E-mail: kongyf@nankai.edu.cn (Y. Kong)

## Abstract

Lithium niobate (LiNbO₃, LN) plays an important role in holographic storage, and molybdenum doped LiNbO₃ (LN:Mo) is an excellent candidate for holographic data storage. In this paper, the basic features of Mo doped LiNbO₃, such as the site preference, electronic structure, and the lattice distortions, have been explored from first-principles calculations. Mo substituting Nb with its highest charge state of +6 is found to be the most stable point defect form. The energy levels formed by Mo with different charge states are distributed in the band gap, which are responsible for the absorption in the visible region. The transition of Mo in different charge states implies molybdenum can serve as a photorefractive center in LN:Mo. In addition, the interactions between Mo and intrinsic or extrinsic point defects are also investigated in this work. Intrinsic defects $V_{Li}^{-}$ could cause the movement of the $Mo_{Nb}^{+}$ energy levels. The exploration of Mo, Mg co-doped LiNbO₃ reveals that although Mg ion could not shift the energy level of Mo, it can change the distribution of electrons in Mo and Mg co-doped LN (LN:Mo,Mg) which help with the photorefractive phenomenon.

## 1. Introduction

Lithium niobate (LiNbO₃, LN) is one of the most used synthetic crystals as the compound presents fascinating photorefractive characteristics which make it possible for useful devices and platforms for integrated photonics.[1-3] As we know, the photorefractive properties of LiNbO₃ can be improved by doping with $Fe^{3+/2+}$, $Mn^{3+/2+}$, $Cu^{2+/+}$, or $Bi^{3+/2+}$ ions.[4-9] These materials are promising candidates for holographic storage applications. For example, bismuth (Bi) and magnesium (Mg) co-doped LiNbO₃ (LN:Bi,Mg) crystals can be used in the dynamic holographic display.[5,6] While, molybdenum-doped LiNbO₃ (LN:Mo) is the only possible one for holographic storage from the ultraviolet to the visible with considerably shorter response time.[10,11] However, there are still not enough accurate and detailed knowledge about LN defects in experiments. Due to the complicated structures of the LN crystals, it is difficult to trace the relationship between these different characteristics and different dopants directly by experimental techniques. The reason for the distinguished performance of LN:Mo is still unknown.

Recently, theoretical investigations play a more and more important role in the exploring the properties of crystals and explaining experimental phenomena. A number of theoretical simulations of the pure and doped LiNbO₃ have been carried out in the past few years.[12-16] In 2010, Xu *et al.* explored the site selection of $Fe^{2+/3+}$ ions,[12] and Li *et al.* revealed the relationship between the electronic states of Fe ions and the light absorption in the visible region.[13] In the case of bismuth-doped LN, the founding of special lone electron pair effect and small bound electron polaron are helpful in explaining the improved diffraction efficiency of Bi doped LN.[14,15] However, there is no related report about molybdenum-doped $LiNbO_3$. The systematic analysis of the theoretical calculations on currently used dopants reveals that their charge states are all below +5, the valence of Nb, and the results show that these dopants preferably occupy the Li sites. Even in the vanadium-doped $LiNbO_3$, the V is found to prefer to substitute Li at its highest charge state of +5.[16] It is known that the highest charge state of Mo is +6, which is higher than the charge state of Nb ions. Whether Mo-doped $LiNbO_3$ will occupy Nb sites or not is a tempting problem to be explored. And, the interaction between the Mo and intrinsic defects is another important characteristic that should be understood as the intrinsic defects $Nb_{Li}^{4+}$ and $V_{Li}^{-}$ are inevitable in LN crystals. The distribution of these point defects and the interaction between them are closely related to the properties of the crystal.

In addition, co-doping non-photorefractive ions with photorefractive ions could improve the laser-induced optical damage resistance and the response speed of $LiNbO_3$ crystals.[17-19] For example, the response time of Mo, Mg co-doped $LiNbO_3$ (LN:Mo,Mg) is dramatically shortened compared with LN:Mo.[11] The reason for the more excellent properties of Mo co-doped with Mg, In and Zr is hard to be figured out from the experiments. Therefore, theoretical calculation provides another way to understand the relationship between Mo and these non-photorefractive ions which will help us to select proper dopants and proper concentration in the crystal growth process.

In this work, based on the density functional theory (DFT), the defect formation energies, lattice distortions and electronic properties of LN:Mo point defects were explored.[20,21] The combination of experimental study and theoretical investigation on the site selection of Mo and its charge state are carried out. In the present calculations, we intend to reveal the distributions and interactions between Mo and the intrinsic $Nb_{Li}^{4+}$ and $V_{Li}^{-}$ and extrinsic point defects. Furthermore, the density of states (DOS) and charge difference map of LN:Mo and LN:Mo,Mg are calculated to elucidate the effect of Mg co-doping on the electron distribution. Calculation details are collected in the Method part.

## 2. Method

We used the Vienna *ab initio* simulation package (VASP)[22,23] which performs an iterative solution of the Kohn−Sham equations with a plane-wave basis set. The energy cutoff for plane waves was 400 eV. The electron interactions for atoms were described by the projector-augmented wave (PAW) method developed by Blöchl with the Perdew, Burke, and Ernzerhof (PBE) approximation for the exchange and correlation.[24,25] Therefore, two outer electrons of Li ($2s^1$), six of O ($2s^2$, $2p^4$), eleven of Nb ($4p^6$, $4d^4$, $5s^1$), and six of Mo ($4d^5$, $5s^1$) were explicitly treated. Mo substituting Li ($Mo_{Li}$) and Mo substituting Nb site ($Mo_{Nb}$) point defects are calculated in the 240-atom hexagonal supercell. The same supercell is implemented in the situation that Mo$_{Li}$ and Mo$_{Nb}$ point defects coexist with intrinsic defects or extrinsic point defects Mg substituting Li site (Mg$_{Li}$), respectively. For LiNbO$_3$ hexagonal supercell, the length of the c-axis is more than twice the length of the a and b axis, thereby, a 4×4×2 k-points mesh over the Brillouin zone with half number of k points in the z direction (c axis) as in x and y directions generated by the Monkhost-Pack scheme is employed.[26] As the supercell we employed is big enough, and to minimize the computational cost and save time, the Monkhost-Pack scheme used for defect pair calculation is a 2×2×1 k-points mesh. For all the calculations, the structure is optimized with a force convergence criterion of 0.01 eV/Å.

Defect formation energy (DFE) as a criterion for judging the stability of point defects and defect clusters. The lower formation energy corresponds to the more stable defects.[27] The DFEs of Mo$_{Li}$ and Mo$_{Nb}$ point defects are calculated to explore the most stable charge state of the two point defects in LN:Mo. In order to find out the most suitable location distribution of Mo$_{Li}$ and Mo$_{Nb}$ point defects with intrinsic point defects lithium vacancy and Nb antisite, the DFEs of defect pairs are calculated, too. In general, DFE (E$_f$) can be calculated by [28,29]

$$E_f(X^q) = E^{total}(X^q) - E^{total}(\text{perfect}) + \sum_i n_i \mu_i + q(E_F + E_v + \Delta V) \quad (1)$$

where $X$ represents the point defect or the defect pairs and can be charged with $q$ or electrically neutral. $E^{total}(X^q)$ is the total energy of the bulk with defect $X$, while $E^{total}$(perfect) is the total energy of the pristine supercell. $i$ is the species of atoms that have been added to or removed from pristine crystal, $n_i$ is the number of atoms $i$, and $\mu_i$ indicates the chemical potential of corresponding atoms $i$. $\Delta \mu_i$ is defined as differences from the bulk values of the chemical of atoms $i$. The chemical potential of Nb ($\mu_{Nb}$), Li ($\mu_{Li}$) and O ($\mu_O$) atoms are calculated with DFT-PBE functional. They depend on the preparation conditions, and vary with the change of different reference phase in the constraints range. $\mu_{Nb}$, $\mu_{Li}$ and $\mu_O$ should also satisfy with the formation of enthalpy of their oxides Li$_2$O and Nb$_2$O$_5$,[28,30]

$$2\Delta\mu(\text{Li}) + \Delta\mu(\text{O}) = -\Delta H_f^{Li_2O} \quad (2)$$

$$2\Delta\mu(\text{Nb}) + 5\Delta\mu(\text{O}) = -\Delta H_f^{Nb_2O_5} \quad (3)$$

The relationship of $\mu_{Nb}$, $\mu_{Li}$ and $\mu_O$ are also constrained by the equation of forming the stable LiNbO$_3$:

$$\Delta\mu(\text{Li}) + \Delta\mu(\text{Nb}) + 3\Delta\mu(\text{O}) = -\Delta H_f^{LiNbO_3} \quad (4)$$

In addition, the restrictions of $\mu_{Mo}$, is according to the experimental condition that Mo ions is from its oxide MoO$_3$, therefore, $\mu_{Mo}$ should satisfy with the requirement of forming the oxide MoO$_3$,[10]

$$\Delta\mu(\text{Mo}) + 3\Delta\mu(\text{O}) = -\Delta H_f^{MoO_3} \quad (5)$$

Similarly, the chemical potentials of extrinsic defects Mg ions $\mu_{Mg}$ is determined according to the formation enthalpy of their oxides,

$$\Delta\mu(Mg) + \Delta\mu(O) = -\Delta H_f^{MgO} \quad (6)$$

We have plotted the thermodynamically stable region of the LN in the former work to define the chemical potential of its components.[31] Table 1 lists the chemical potentials of Li, Nb, O, and Mo under Li-rich and Li-deficient conditions. As the as-grown crystals and films are Li-deficient composition, the chemical potentials of Nb and Li are employed according to the Li-deficient condition. The chemical potentials of Mo and Mg which are non-lithium niobate components are also calculated based on the chemical potential of O under the Li-deficient conditions. $E_V$ is the valence band maximum (VBM) of crystals and $E_F$ is the Fermi level in regard to the VBM. $\Delta V$ aligns with the reference potential difference between the defect supercell and the pristine crystal and it is related to the volume of the supercell. This term can be obtained from the electrostatic potentials difference between the region of defect and the region far from the defect.[32] The value of $\Delta V$ is 0.18 eV to the maximum in this calculation, and it brings a small impact when it plus with charge q.

As the total internal energies obtained from DFT calculations correspond to the Helmholtz free energy at zero temperature, there is a free energy correction between VASP work environment and real condition. The electronic entropy is negligible due to the large band gap of LiNbO$_3$, nonetheless, strain effects can be considered negligible in a large cell. The free energy ($F = E - TS$) is mainly related to the configuration contribution of point defects and defect clusters. At the room temperature of 300 K, the entropy of point defect and defect pair is about 0.16-0.20 eV.[33-36]

Table 1. Chemical potentials of components in LN and doped LN by DFT-PBE under Li-rich and Li-deficient conditions.

| Component | Chemical potential (eV) | |
|---|---|---|
| | Li-rich | Li-deficient |
| Li | -2.56 | -3.63 |
| Nb | -19.92 | -18.96 |
| O | -5.63 | -5.59 |
| Mo | -18.71 | -18.83 |
| Mg | -6.34 | -6.30 |

Binding energy $E_b$ as a criterion for judging the stability of a defect pair $X_1X_2$, usually be defined in terms of the formation energies[16,28]

$$E_b[(X_1X_2)^q] = E_f[(X_1X_2)^q] - E_f[(X_1)^{q_1}] - E_f[(X_2)^{q_2}] \quad (7)$$

where $q = q_1 + q_2$, the negative binding energy means that the energy required to separate the defect pair into two individual defects $X_1$ and $X_2$ is more than the formation energy of defect pair $X_1X_2$, which indicates a stable defect pair.

## 3. Results and Discussion
### 3.1. Point defects in the LN:Mo crystal

In this study, we first calculate the formation energies of Mo substituting Li ($Mo_{Li}$) and Nb ($Mo_{Nb}$) at all possible charge states in a 240-atom supercell based on the DFT-PBE. The results are shown in Figure 1. It can be seen that Mo prefers to occupy the Li site at the +6 charge state when the Fermi energy is close to the VBM. With the increase of Fermi energy which is mainly due to the increase of dopants concentration, the $Mo_{Li}^{5+}$ transfers to $Mo_{Li}^{4+}$ when the Fermi energy $E_F = 1.50$ eV. At this time, the charge state of Mo and Nb are the same, and then $Mo_{Li}^{4+}$ transfers to $Mo_{Li}^{2+}$ directly by capturing two electrons, it seems like the transition behavior of $Nb_{Li}$. $Mo_{Li}^{+}$ presents as the most stable charge state in a short range of Fermi energy from 1.95 eV to 2.08 eV. The results indicate that $Mo_{Li}^{3+}$ is metastable, due to the negative $U$ effect,[31,37] the thermodynamic transition level $\varepsilon$ (+4/+3) is higher than $\varepsilon$ (+3/+2), therefore, the $Mo_{Li}^{3+}$ cannot be shown as a stable charge state in Figure 1. While in the most LiNbO$_3$ crystals, the Fermi level lies in the lower half of the band gap, therefore, $Mo_{Li}^{5+}$, $Mo_{Li}^{4+}$, and $Mo_{Li}^{2+}$ are stable point defects of Mo in Li sites. Compared with the results of $Nb_{Li}$ point defect, we found that $Mo_{Li}$ possesses a higher formation energy in the whole Fermi energy range. It is difficult for Mo ions to push the Nb-antisite to the normal position like other dopants.

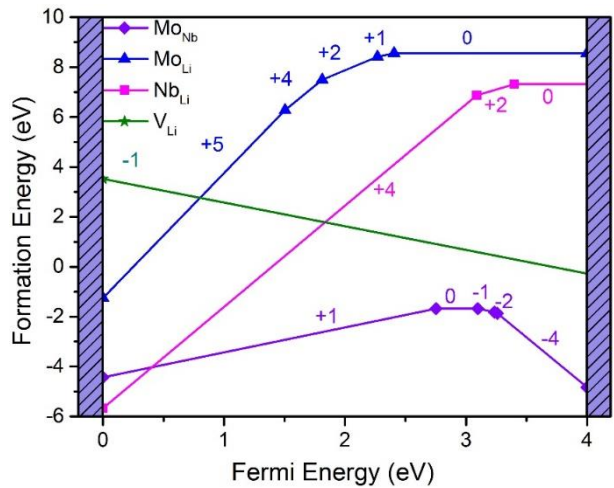

Figure 1. Defect formation energies of point defects $Mo_{Li}$ and $Mo_{Nb}$ with the most stable charge state, as well as the intrinsic defects $Nb_{Li}$ and $V_{Li}$ as a function of the Fermi energy. The Fermi energy range is from VBM to CBM.

In the case of $Mo_{Nb}$, $Mo^{6+}$ is the most stable charge state in the most possible Fermi energy range, then $Mo^{6+}$ will transfer to $Mo^{5+}$ which presents the same charge state as Nb ions. $Mo^{5+}$ as the most stable state is maintained within the Fermi range from 2.75–3.10 eV. Mo substituting Nb ions with +4, +3 also appears in a short range. When the Fermi energy close to the conduction band minimum (CBM), the most stable charge state of $Mo_{Nb}$ is -4. Same as Mo occupies the Li site, the lowest charge state of Mo ions at Nb site occurs near the CBM, which indicates a high concentration of dopants. In the entire Fermi energy range, $Mo_{Nb}$ possesses lower formation energy than $Mo_{Li}$, which illustrates that Mo ions prefer to substitute Nb ions to form $Mo_{Nb}^{+}$ point defect. Single crystal X-ray diffraction experiment results also confirmed that Mo occupies the niobium site in the crystals.[13] As the fact that in the most LiNbO$_3$ crystals, the Fermi level lies in the lower half of the band gap, Mo ions with higher charge state +4, +5, +6 are more reasonable. The experiment results of X-ray photoelectron spectroscopy (XPS) imply the coexistence of $Mo^{6+}$, $Mo^{5+}$, and $Mo^{4+}$.[10]

Comparing the formation energy of $Mo_{Nb}$ to the stable intrinsic $Nb_{Li}^{4+}$ and $V_{Li}^{-}$ defects, we found that the formation energy of $Mo_{Nb}$ is much lower than that of $V_{Li}$ in the possible Fermi range. And $Mo_{Nb}$ is easier to form beyond $E_F = 0.4$ eV compared with $Nb_{Li}$, while $Nb_{Li}$ is easier below $E_F = 0.4$ eV. The results indicate that $Mo_{Nb}$ and $Nb_{Li}^{4+}$ can coexist in the crystals, therefore, $Mo_{Nb}^{+}$ is formed easily in both stoichiometric and congruent LiNbO3 crystals. We will discuss the interaction between them later.

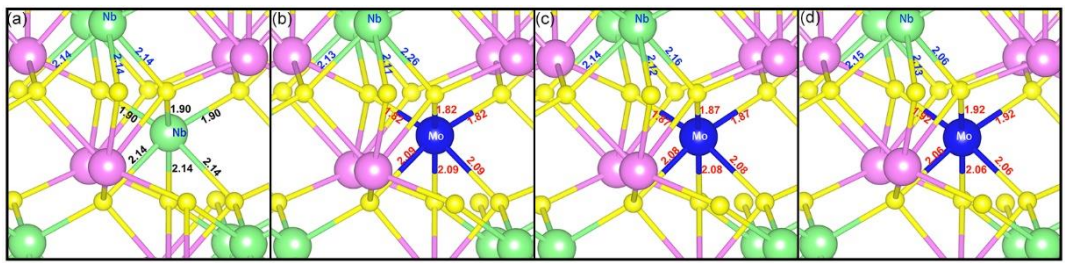

Figure 2. The local lattice of distortion of (a) pristine and (b) $Mo_{Nb}^{+}$, (c) $Mo_{Nb}^{0}$, (d) $Mo_{Nb}^{-}$ point defects. The number labeled on the bond is the distance between the defects and its neighboring O ions (the black and red numbers), and the distance between the normal Nb site and its neighboring O ions (the blue numbers). The unit of these numbers is Å.

In order to understand the effect that Mo substituting Nb site bring to the bulk, we investigate the localized structural relaxations of the most stable $Mo_{Nb}^{+}$, $Mo_{Nb}^{0}$ and $Mo_{Nb}^{-}$ point defects in Figure 2. For $Mo_{Nb}^{+}$, as shown in Figure 2(a) and (b), we can see that the Mo-O$_{upper}$ bond length decreases to 1.82 Å compared with the pristine Nb-O bond length 1.90 Å, the Mo-O$_{lower}$ bond length also decreases by about 2.3%. There is a significant shrinkage of the oxygen octahedron around the $Mo_{Nb}$ defect. While in the case of normal Nb site, it can be seen that the distances between Nb and two lower neighboring O site are also a little decrease. However, the O ion that is simultaneously bonded to the normal Nb and Mo ion become an exception, the bond length between the O ion and normal Nb ion increases. The reason for this phenomenon is the stronger covalent bond of Mo-O as compared to Nb-O. And it may lead to a deformation of the oxygen ion electron cloud, which is related to the narrower band gap. This corresponds to the red shift of the absorption edge of the LN:Mo crystals in contrast to CLN.[10] And we can see the similar shrinkage of the oxygen octahedron around the $Mo_{Nb}$ defect in the situation of $Mo_{Nb}^{0}$ and $Mo_{Nb}^{-}$ point defects. While the severity of shrinkage decreases with the decreasing charge state of Mo ions due to the stronger Coulomb repulsion between the trapped electrons and electrons around O ions.

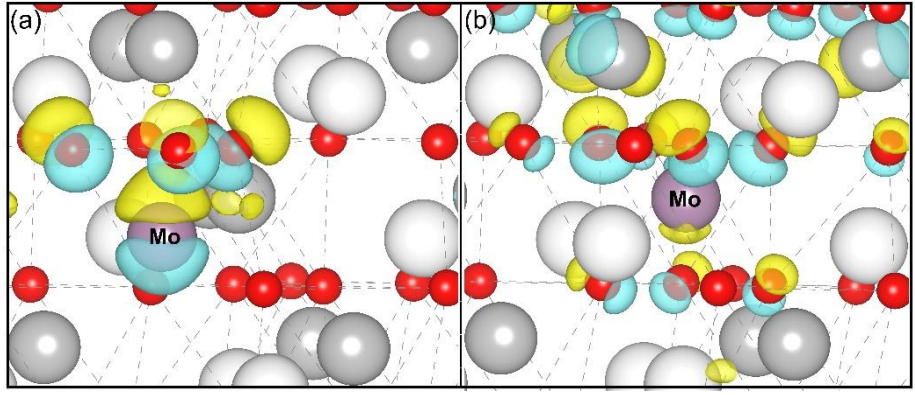

Figure 3. The charge density difference map of (a) $Mo_{Li}^{4+}$ and $Mo_{Li}^{2+}$, (b) $Mo_{Nb}^{+}$ and $Mo_{Nb}^{-}$. The white, gray, red and violet balls denote the Li, Nb, O and Mo atoms respectively.

The discussion about the lattice distortion of point defects lays the foundation for the study of whether Mo can become a photorefractive center or not. As we mentioned above, the transition between $Mo_{Li}^{4+}$ and $Mo_{Li}^{2+}$ is similar to the situation of $Nb_{Li}^{4+}$ and $Nb_{Li}^{2+}$. If $Nb_{Li}^{4+}$ capture two electrons to form $Nb_{Li}^{2+}$, and introduce the lattice distortion between $Nb_{Li}^{2+}$ and its neighbor normal Nb site, then $Nb_{Li}^{2+}$ can be treated as a $4d^1$-$4d^1$ bipolaron.[38] In Figure 3(a), we show the charge density difference map of $Mo_{Li}^{4+}$ and $Mo_{Li}^{2+}$. It can be seen that the two captured electrons are distributed around the $Mo_{Li}^{2+}$ and its neighboring Nb. The results are in contrast to the situation of $Nb_{Li}^{2+}$, therefore, $Mo_{Li}^{2+}$ can also be treated as a $4d^1$-$4d^1$ bipolaron, which means that $Mo_{Li}^{2+/4+}$ serves as a photorefractive center in the crystals. Whether Mo substituting Nb site can also be the photorefractive center is a problem to be discussed. If $Mo_{Nb}^{+}$ capture two electrons then it can be treated as the $Mo_{Nb}^{-}$, another stable charge state in Mo point defect. In Figure 3(b), the charge density difference map of $Mo_{Nb}^{+}$ and $Mo_{Nb}^{-}$ is plotted. Seen from it, the trapped electrons are found also mainly gathered around $Mo_{Nb}^{-}$ as $Mo_{Li}^{2+}$ which leads to the lattice distortion we discussed above. Similar results illustrate that Mo$_{Nb}$ may also be a photorefractive center. In LN:Mo, both Mo$_{Li}$ and Mo$_{Nb}$ are contributors to the photorefraction effect.

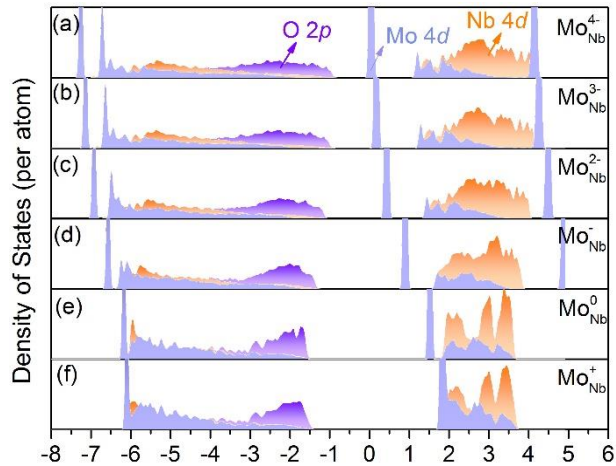

Figure 4. PDOS of the Mo$_{Nb}$ point defect with charge state from -4 to +1 (a)-(f). The contribution of LiNbO$_3$ elements are shown with different colors. The contribution of Mo$_{Nb}$ is labeled as light blue.

Differently from the dopants such as vanadium and bismuth, Mo substituting Nb position may present a positive charge state. To explore the particularity of $Mo_{Nb}$, we draw the partial density of states (PDOS) of $Mo_{Nb}$ point defects with different charge states in Figure 4. Seen from it, there is a density of states peaks which are made up of Mo-4$d$ electrons in the middle of the band gap of LN:Mo with $Mo_{Nb}$ point defects. It illustrates that $Mo_{Nb}$ forms new energy levels in the band gap. With Mo ion losing electrons, the energy level of Mo gradually approaches to CBM, and when all the six outer electrons of Mo are lost, the contribution of $Mo_{Nb}$ overlap to the contribution of Nb elements. The plentiful energy levels facilitate the electronic transitions between different energy levels, which corresponding to the experiment results that there are absorption peaks in the visible range.[10,11] Therefore, Mo-doped $LiNbO_3$ indeed improves the photorefractive effect of crystals in the all visible light range.

### 3.2. Mo point defects with intrinsic point defect

From the analysis of the defect formation energies of the most stable charge state $Mo_{Nb}^{+}$ as well as the intrinsic defects $Nb_{Li}^{4+}$ and $V_{Li}^{-}$, we believe that $Mo_{Nb}^{+}$, $Nb_{Li}^{4+}$ and $V_{Li}^{-}$ coexist in the $LiNbO_3$ crystal. Based on the above result, the distribution of $Mo_{Nb}^{+}$ and $Nb_{Li}^{4+}$ defects is explored in Figure 5(a). As both of them present the positive charge state, when they are close to each other, the distance between them would be enlarged due to the obvious Coulombic repulsion. The binding energy of the $Mo_{Nb}^{+}$ and $Nb_{Li}^{4+}$ defect pair increases with the distance increased from 1NN to 3NN, then it decreases to -1.5 eV until the $Nb_{Li}^{4+}$ is far away from the $Mo_{Nb}^{+}$. The binding energies of $Mo_{Nb}^{+}$ and $Nb_{Li}^{4+}$ defect pair are found to be positive except when they separate from each other, therefore, it is suitable to separate $Mo_{Nb}^{+}$ and $Nb_{Li}^{4+}$ to two single defects. While as Mo substitute Nb site, the extra Nb may lead to more $Nb_{Li}^{4+}$ antisite. On the one hand, the additional $Nb_{Li}^{4+}$ can also play the role of the photorefractive center; on the other hand, to maintain an electrically neutral environment in the crystal, more lithium vacancies are required for charge compensation. The distribution of $Mo_{Nb}^{+}$ and $V_{Li}^{-}$ is also explored in Figure 5(b).

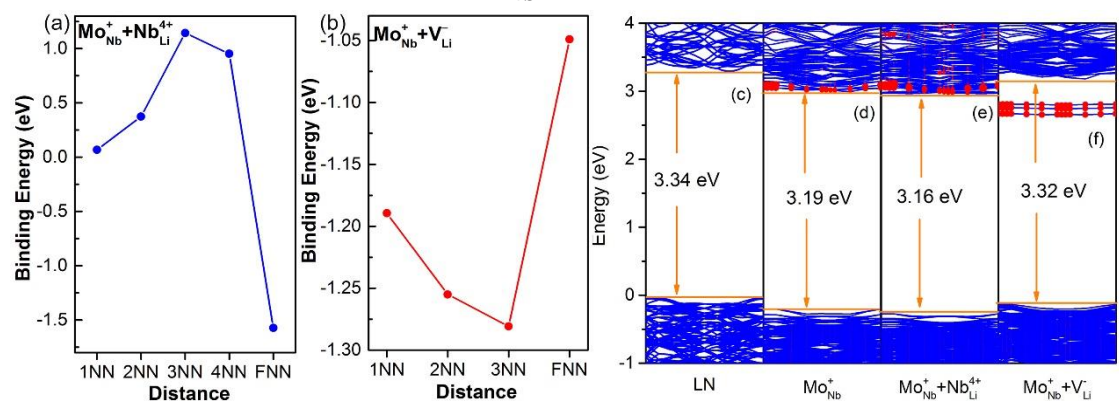

Figure 5. The binding energy of $Mo_{Nb}^{+}$ defect with intrinsic defect (a) $Nb_{Li}^{4+}$ and (b) $V_{Li}^{-}$ with different distances. 1NN to 4NN are corresponding to the position of $Nb_{Li}^{4+}$ or $V_{Li}^{-}$ which lie in the first to the forth nearest neighboring sites of $Mo_{Nb}^{+}$, and the FNN is a site far away from the $Mo_{Nb}^{+}$ site. (c)-(f) show the band structures of LN pristine and LN with $Mo_{Nb}^{+}$, $Mo_{Nb}^{+}+Nb_{Li}^{4+}$, and $Mo_{Nb}^{+}+V_{Li}^{-}$ defects. The red lines are the energy levels of Mo ions.

The binding energy of $Mo_{Nb}^+$ and $V_{Li}^-$ defect pair shows the lowest formation energy when $V_{Li}^-$ is the third nearest neighbor of $Mo_{Nb}^+$. And the highest binding energy happened in the situation that $V_{Li}^-$ is far away from $Mo_{Nb}^+$. The results are different from the case of $Mo_{Nb}^+ + Nb_{Li}^{4+}$ defect pair. It illustrates that $Mo_{Nb}^+ + V_{Li}^-$ is a stable defect pair with a small distance with each other. As we know that $V_{Li}^-$ will affect the oxygen ions around it, by the way of $V_{Li}^-$, $Mo_{Nb}^+$ may introduce the effect on the distribution of electrons around them.

In Figure 5(c)-(f), we show the band structures of LiNbO3 bulk, $Mo_{Nb}^+$ point defect, $Mo_{Nb}^+ + Nb_{Li}^{4+}$ and $Mo_{Nb}^+ + V_{Li}^-$ defect pairs. We can see a narrower band gap with Mo ions added to the LN system. This is in line with our previous conclusion that the Mo-O bond is more stable than the Nb-O bond which may lead to the red shift of the absorption edge. Seen from Figure 5(d), there is no obvious energy level in the band gap, the contribution of Mo ions distributed in the conduction band, and they overlap with the conduction band. It indicates that Mo ions can make an extra absorption in the ultraviolet region when the electrons transit from the valence band to the Mo energy level. The results are matching with the experiments that there is an absorption peak at 337 nm. The band structure of $Mo_{Nb}^+ + Nb_{Li}^{4+}$ defect pair is shown in Figure 5(e). There are no significant differences between the $Mo_{Nb}^+$ point defect and $Mo_{Nb}^+ + Nb_{Li}^{4+}$ defect pair. This performance also provides proof that $Mo_{Nb}^+$ and $Nb_{Li}^{4+}$ have no interaction with each other, which indicate that $Mo_{Nb}^+$ and $Nb_{Li}^{4+}$ should be separated from each other. In the band structure of $Mo_{Nb}^+ + V_{Li}^-$ defect pair, there are some new energy levels in the band gap, which are about 0.4 eV away from the bottom of the conduction band. These defect levels provide an opportunity of electron transition from defect levels to conduction band. It may correspond to the absorption peak in the visible light region.

**3.3 Mo point defect with extrinsic point defect**

In this section, we investigated the properties of LN:Mo,Mg. Learning from the result of Li et al, Mg ions prefer the Li site in the LN crystals.[13] Therefore, when Mo co-doped with Mg ions, Mo ions will occupy Nb ions while Mg ions will substitute Li site. The distribution of the Mo ions with Mg ions in different distance is discussed in Figure 6, respectively. $Mg_{Li}^+$ as the first, second, third and faraway nearest neighbors of $Mo_{Nb}^+$ point defect are considered, respectively.

Seen from Figure 6, as the function of the Coulombic repulsion between two positive point defects, when $Mg_{Li}^+$ is far away from $Mo_{Nb}^+$, the defect pairs show the lowest binding energy. In Figure 6(a), the negative binding energies illustrate that it is easy to form a defect pair of $Mg_{Li}^+ + Mo_{Nb}^+$. To maintain the neutral environment in the crystal, $Mg_{Li}^+ + Mo_{Nb}^+$ defect pair should be charge compensated by the intrinsic defect $V_{Li}^-$ to form a stable defect cluster. However, if $Mg_{Li}^+$ set in the faraway site of $Mo_{Nb}^+$, the interaction between them is too weak to be explored. Therefore, $Mg_{Li}^+$ as the first nearest neighbor is also taken into consideration in the process of clustering. Figure 6(b) shows the structure of $Mg_{Li}^+ + Mo_{Nb}^+$ defect pair, the distribution of $V_{Li}^-$ are discussed in the following.

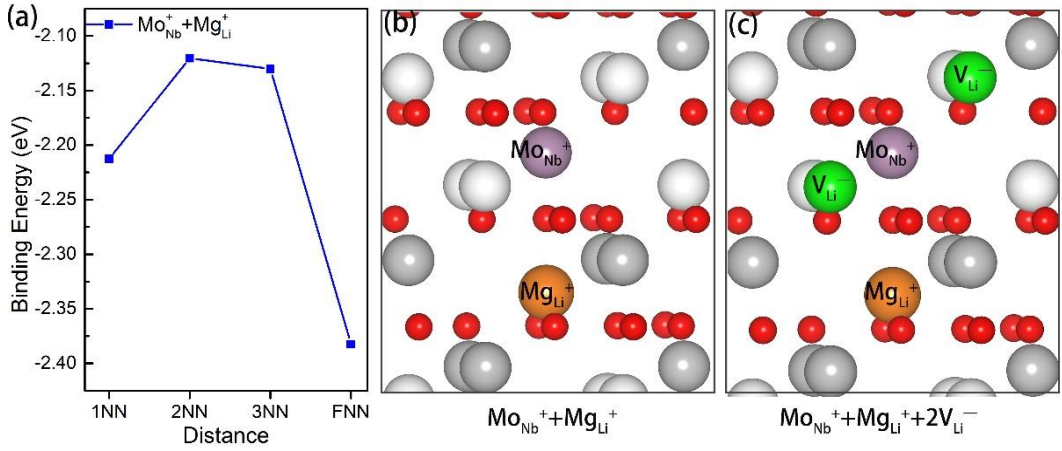

Figure 6. (a) The binding energy of $Mo_{Nb}^+$ defect with extrinsic defect $Mg_{Li}^+$, 1NN to 3NN are corresponding to the position of $Mg_{Li}^+$ which lie in the first to the third nearest neighboring sites of $Mo_{Nb}^+$, the FNN is a site far away from $Mo_{Nb}^+$ site. (b) the structure of $Mg_{Li}^+ + Mo_{Nb}^+$ defect pair, (c) the distribution of $V_{Li}^-$ in the $Mg_{Li}^+ + Mo_{Nb}^+ + 2V_{Li}^-$ cluster.

Table 2. The formation energies of different lithium vacancies distributions around $Mg_{Li}^+ + Mo_{Nb}^+$ defect pair with a reference of $Mo_{Nb}^+$.

| Distribution of the two lithium vacancies | Formation energies (eV) |
| --- | --- |
| 2NN+2NN | 1.42 |
| 3NN+3NN | 1.49 |
| 2NN+3NN | 1.37 |

According to the results in Table 2, we found the structure that when one of the lithium vacancies is the second nearest neighbor of $Mo_{Nb}^+$ while the other one is the third nearest neighbor of $Mo_{Nb}^+$ is the most stable cluster of $Mg_{Li}^+ + Mo_{Nb}^+ + 2V_{Li}^-$. Then, one lithium vacancy is upper the defect pair $Mg_{Li}^+ + Mo_{Nb}^+$, the other is in the defect pair as shown in Figure 6 (c). The result is consistent with the above conclusion that $V_{Li}^-$ is prefer to be the 3NN of $Mo_{Nb}^+$. In order to explore the relationship between $Mo_{Nb}^+$ and $Mg_{Li}^+$, we draw the PDOS of $Mg_{Li}^+ + Mo_{Nb}^+$ defect pair, and $Mo_{Nb}^+$ point defect, $Mo_{Nb}^+ + V_{Li}^-$ defect pair, $Mg_{Li}^+ + Mo_{Nb}^+ + 2V_{Li}^-$ defect cluster for comparison. From the PDOS of $Mo_{Nb}^+$ in Figure 7(a), it can be seen that the contribution of Mo 4$d$ overlaps with the states of Nb 4$d$. From the band structure in Figure 5(e) and (f), under the influence of $V_{Li}^-$, the energy levels of Mo are separated from the conduction band. The same conclusion can also be obtained from Figure 7(b). While the combination with $Mg_{Li}^+$ point defect does not change the distribution of Mo 4$d$ features, the $d$ electrons of Mo and Nb still overlap together. When $Mg_{Li}^+ + Mo_{Nb}^+$ defect pair forming the stable defect cluster with $V_{Li}^-$, the states of Mo 4$d$ separate from the conduction band too, however, the Mo 4$d$ electrons are a little closer to the conduction band compared to the situation of $Mo_{Nb}^+ + V_{Li}^-$ defect pair. The properties of $Mo_{Nb}^+$ have changed under the combined action of $Mg_{Li}^+$ and $V_{Li}^-$. To seek for the effect of Mg ions, the electronic charge difference map between $Mg_{Li}^+ + Mo_{Nb}^+$ defect pair and $Mo_{Nb}^+$ point pair are given in Figure 7(e). There is a significant loss of electronics around the Mg ion. And the distribution of electrons of the O atoms that around the Mg ion also change a lot. It illustrates that the oxygen atoms are more active which will help with the decrease of response time.[5,39] Therefore, Mo and Mg co-doped LiNbO$_3$ possess more superior photorefractive properties compared with Mo doped LiNbO$_3$.

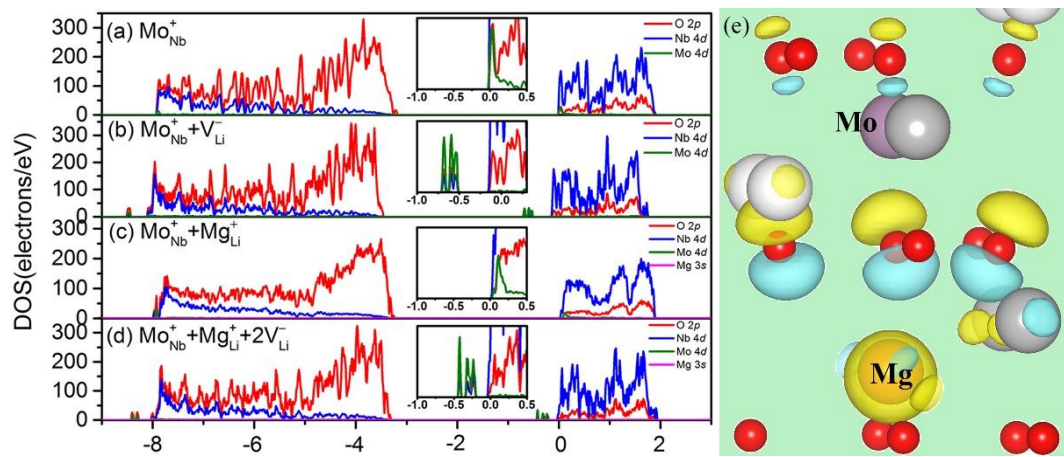

Figure 7. The density of states of (a) $Mo^+_{Nb}$ point defect, (b) $Mo^+_{Nb}+V^-_{Li}$ defect pair, (c) $Mg^+_{Li}+Mo^+_{Nb}$ defect pair, and (d) $Mg^+_{Li}+Mo^+_{Nb}+2V^-_{Li}$ defect cluster. The magnified area shows that the influence of Mo atoms. (e) Electronic charge difference map between $Mg^+_{Li}+Mo^+_{Nb}$ and $Mo^+_{Nb}$. Yellow and blue ellipsoid represent electron depletion and accumulation, respectively. The white, gray, red, orange and purple balls denote the Li, Nb, O, Mg and Mo atoms, respectively.

## 4. Conclusion

In summary, we have performed the detailed first-principles calculations of the possible site preference in isolate defects and corresponding charge state in LN:Mo. Mo substituting Nb with its highest charge state +6 is found to be the most stable point defect form. Both Mo$_{Li}$ and Mo$_{Nb}$ are photorefractive centers to improvement its photorefractive performance in LiNbO$_3$ crystals. The distribution of defect energies from VBM to CBM formed by Mo ions with different charge states is in line with the absorption in the visible region. While the energy levels of $Mo^+_{Nb}$ in the conduction band are responsible for the absorption in the ultraviolet region. Also, $Mo^+_{Nb}$ and $Nb^{4+}_{Li}$ point defects are found should separate from each other, and they show a weak influence on each other. While the $Mo^+_{Nb} + V^-_{Li}$ defect pair is stable when $Mo^+_{Nb}$ and $V^-_{Li}$ are the third nearest neighbors of each other, and the $V^-_{Li}$ could cause the movement of the $Mo^+_{Nb}$ energy levels. In addition, the combination with $Mg^+_{Li}$ point defect do not shift the energy level of Mo, and Mg ions have an influence on the distribution of electrons of LN:Mo,Mg, then the photorefractive properties are improved.

## Acknowledgment

We gratefully acknowledge financial support from National Natural Science Foundation of China with grant number [11674179] and [61705116], and the Program for Changjiang Scholars and the Innovative Research Team in University with grant number [IRT_13R29].